\documentclass{elsart4-2}

\usepackage{hyperref}
\usepackage{graphicx}

\usepackage[english,francais]{babel}
\usepackage[latin1]{inputenc} 

\newtheorem{e-proposition}[theorem]{Proposition}

\newtheorem{e-definition}[theorem]{Definition\rm}

\renewcommand{\theequation}{\arabic{equation}}
\def\og{\leavevmode\raise.3ex\hbox{$\scriptscriptstyle\langle\!\langle$~}}
\def\fg{\leavevmode\raise.3ex\hbox{~$\!\scriptscriptstyle\,\rangle\!\rangle$}}

\newcommand{\DoubleFigureWidth}{100mm}
\newcommand{\SingleFigureWidth}{50mm}

\begin{document}

\begin{frontmatter}


\selectlanguage{english}
\title{The optical calcium frequency standards of PTB and NIST
\thanksref{JAN}
}
\selectlanguage{francais}
\title{Les étalons de fréquence optique au calcium de la PTB et du NIST}
\thanks[JAN]{Dedicated to J.~Hall on the occasion of his 70th birthday.\\
           Work of an agency of the {US} government; not subject to copyright.}


\selectlanguage{english}
\author[PTB]{U. Sterr}
\ead{uwe.sterr@ptb.de}
\author[PTB]{C. Degenhardt}
\author[IQO]{H. Stoehr}
\author[IQO]{Ch. Lisdat}
\author[PTB]{H. Schnatz}
\author[PTB]{J. Helmcke}
\author[PTB]{F. Riehle}
\author[NIST]{G. Wilpers}
\author[NIST]{Ch. Oates}
\ead{oates@boulder.nist.gov}
\author[NIST]{L. Hollberg}

\address[PTB]{Physikalisch-Technische Bundesanstalt, Bundesallee 100, 38116 Braunschweig, Germany}
\address[IQO]{Institut f\"ur Quantenoptik, Universit\"at Hannover, Welfengarten 1, 30167 Hannover, Germany}
\address[NIST]{National Institute of Standards and Technology, 325 Broadway, Boulder Co. 80305, USA}

\begin{abstract}
We describe the current status of the Ca optical frequency standards with laser-cooled neutral atoms realized in two different laboratories for the purpose of developing a possible future optical atomic clock.
Frequency measurements performed at the Physikalisch-Technische Bundesanstalt (PTB) and the National Institute of Standards and Technology (NIST) make the frequency of the clock transition of $^{40}$Ca one of the best known optical frequencies (relative uncertainty $1.2\cdot10^{-14}$) and the measurements of this frequency in both laboratories agree to well within their respective uncertainties.
Prospects for improvement by orders of magnitude in the relative uncertainty of the standard look feasible.
%

\vskip 0.5\baselineskip

\selectlanguage{francais}
\noindent{\bf R\'esum\'e} \vskip 0.5\baselineskip \noindent Nous
décrivons le présent état de l´art des étalons de fréquence
optique avec des atomes neutres de Ca refroidis par laser,
réalisés dans deux différents laboratoires, dans le but de
développer éventuellement une future horloge atomique dans le
domaine optique. Les mesures de fréquences réalisées à la
{\it{Physikalisch-Technische Bundesanstalt}}
 (PTB) et au {\it{National Institute of Standards and Technology}} (NIST) ont permis d´établir la
 fréquence d´horloge du $^{40}$Ca parmi les fréquences optiques les mieux connues
 (avec une exactitude relative de $1.2\cdot10^{-14}$), ces mesures de fréquence étant en bon accord
 dans les deux laboratoires dans la limite de leur incertitude respective.
 Potentiellement, une amélioration par plusieurs ordres de grandeur de l´exactitude
 relative de l´étalon semble possible.
%

\keyword{Optical frequency standard; atomic clock; laser
spectroscopy} \vskip 0.5\baselineskip \noindent{\small{\it
Mots-cl\'es~:} Etalons de fréquence optique; horloge atomique;
spectroscopie laser}

} 
\end{abstract}

\end{frontmatter}


\selectlanguage{english}

\section{Introduction}
\label{Introduction}
The enormous progress in laser spectroscopy, laser cooling and trapping of ions and neutral atoms, combined with novel efficient methods for precision optical frequency metrology has had a strong impact on the development of high-performance optical frequency standards and optical clocks. Such standards are widely used and serve as the most accurate realizations of the length unit \cite{qui03}, for tests of fundamental theories and for the determination of fundamental constants with increasing accuracy \cite{hub98a,ude01}.
In contrast to today's atomic clocks operating in the microwave region, the optical frequencies, higher by roughly five orders of magnitude, can lead to a corresponding increase in frequency stability. Moreover, higher clock frequencies should lead to reduced fractional systematic shifts in the measured frequency, thus increasing clock accuracy.

However, to realize the full potential of a high-performance optical frequency standard, two critical issues need to be resolved.  First, residual line broadening and shifts caused by the thermal motion of the atoms need to be reduced or eliminated. Second, technical noise sources in the interrogation oscillators and the detection have to be suppressed. Laser-cooling techniques to achieve microkelvin temperatures have been demonstrated both for neutral atoms \cite{raa87,kat99,bin01a} in magneto-optical traps (MOTs) and single ions  \cite{win78,neu78} trapped in the center of radio-frequency (Paul) traps \cite{pau90}.
The former systems probe a large number of atoms ($\sim 10^7$) released from MOTs in  field-free environments, therefore achieving exceptional signal-to-noise ratio (SNR) but suffering from a residual Doppler effect \cite{tre01,wil03}. The single-ion approach provides a Doppler-free environment but suffers from the limitation in SNR.
To combine the advantages of both approaches a new concept has been proposed to trap ensembles of neutral atoms in an array of optical dipole traps. Initial experiments based on this approach are now under way in several laboratories (e.\,g.\  \cite{tak03}).

Neutral atoms as well as single ions have been successfully applied in time and length metrology. Absolute frequency measurements with relative frequency uncertainties on the level of $10^{-14}$ have been performed for neutral Ca  \cite{ude01,hel03} and H atoms \cite{nie00} and for Hg$^{+}$, In$^{+}$, Sr$^{+}$, and Yb$^{+}$ ions \cite{ude01,zan00,ste01a,mar03}.

A recent advance for absolute frequency measurements in the optical domain utilizes optical frequency comb generators based on mode-locked femtosecond lasers \cite{rei99,ram02a}.  Such a system allows one to effectively eliminate the uncertainty contribution of the measurement process. The capability of such systems to compare optical frequencies with a fractional frequency reproducibility in the range of $10^{-19}$ \cite{ma04} has been shown recently.
Hence the application of optical transitions for atomic clocks \cite{did01} becomes feasible on a level that today is only reached by the best atomic clocks operating in the microwave range.

In this article, we present the results from two different realizations of an optical frequency standard based on ensembles of laser-cooled neutral Ca atoms released from magneto-optical traps at PTB, Braunschweig, Germany and NIST, Boulder, Colorado, USA.
The limitation in the achievable accuracy due to the residual Doppler effect has been overcome on a $10^{-15}$ level in  frequency measurements performed at PTB with a new cooling technique \cite{bin01a}. They might be minimized further to well below $10^{-16}$ by use of a new method developed at NIST based on an idea by Trebst et al.\ \cite{tre01}.
Frequency measurements performed in both places lead to the Ca clock transition frequency being known with one of the lowest uncertainties in the optical range and agree well within the respective uncertainty budgets of the two setups.
We will summarize the results achieved for millikelvin as well as microkelvin atomic ensembles utilized in the frequency standards and present the prominent properties of this Ca clock transition that make it particularly suitable for the realization of a future optical clock.

\section{Experiment}
\label{Experiment}
\begin{figure}
\begin{minipage}[t]{\DoubleFigureWidth}
\begin{center}
\includegraphics[width=8cm]{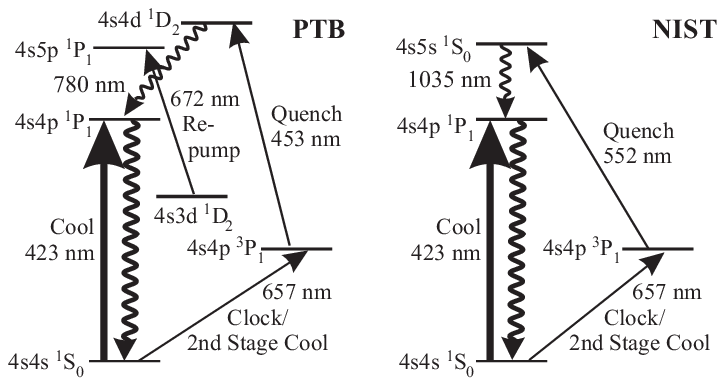}
\caption{Energy level diagram of $^{40}$Ca with the laser transitions relevant for the optical frequency standards at PTB and NIST.}
\label{LevelSchemes}
\end{center}
\end{minipage}
\hfill
\begin{minipage}[t]{\SingleFigureWidth}
\includegraphics[width=\SingleFigureWidth]{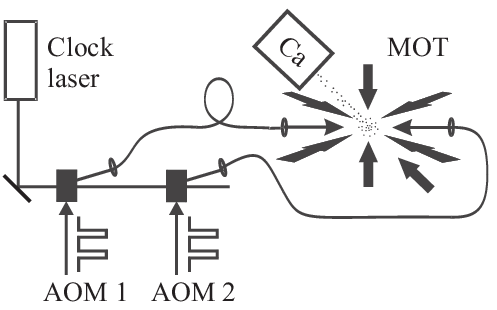}
\caption{Sketch of the MOT and spectroscopy setup.}
\label{Setup}
\end{minipage}
\end{figure}

Both experiments use a continuously repeated measurement sequence consisting of two trapping and cooling stages, manipulation of the cooled atoms for investigating systematics, spectroscopy and detection.  In the first cooling stage atoms are captured from a thermal atomic beam and cooled to millikelvin temperatures by use of a standard MOT based on the broad 423~nm transition (Fig.~\ref{LevelSchemes}) that is almost closed. At PTB, 500~mW of 423~nm light from a frequency-doubled Ti:sapphire laser is used.
A small loss channel with a small fraction of excited atoms on the cooling transition decaying to intermediate long lived states \cite{kur92b}  is closed by use of a repump laser at  672~nm (see Fig.~\ref{LevelSchemes}).
At NIST, 40~mW of frequency-doubled light from a diode laser system is applied, no repump laser is used and a slowing beam is applied to increase the number of atoms in the thermal beam with velocities below 30~m/s that can be trapped by the 423~nm MOT.
In a second cooling stage the atoms are further Doppler cooled in a MOT utilizing the narrow 657~nm clock transition to about 10~$\mu$K. Here the excited state is optically quenched to increase the cooling and trapping forces well above the gravitational force. The two systems use different quenching transitions to increase the photon scattering rates (see Fig.~\ref{LevelSchemes}). From the initial numbers of $8\cdot10^7$ and $5\cdot10^6$ atoms at 3~mK loaded in 350~ms at PTB and 25~ms at NIST up to 40~\% are transferred to the microkelvin ensemble. The times for the second-stage cooling are 20~ms and 7~ms and temperatures of around $12~\mu$K are reached \cite{bin01a,cur03}.
 After the external fields are switched off releasing the atoms from the trap and switching on a homogenous magnetic quantization field, a spectroscopic sequence is applied that excites some fraction of the atoms with laser pulses resonant with the clock transition. Directly after this spectroscopic sequence, the normalized excitation probability of the atomic ensembles is measured. The detection scheme used for this purpose takes advantage of the high number of photons that can be scattered on the 423~nm transition \cite{cur03,oat00,wil02c} and gives a signal that is independent of fluctuations of atom number due to the normalization technique. Thus, the detection does not present a limitation to the principally achievable SNR given by quantum projection noise (QPN) \cite{ita93} (see section~\ref{StabilitySection}).

\subsection{Interrogating the clock transition}

Prior to the interrogation of the clock transition all external fields are switched off to minimize systematic shifts due to electric and magnetic fields, and a homogeneous bias magnetic field of 0.1 -- 0.2 mT is applied. To allow for the eddy currents to decay and the magnetic fields to settle, waiting times of $200~\mu$s with magnetic field coils located in the vacuum are used at PTB, and waiting times of 1~ms are used at NIST with the coils located outside the vacuum chamber.
To probe the clock transition at a resolution close to its natural linewidth, the use of both highly coherent laser radiation and atom interferometry in the time domain is necessary.

\subsubsection{Clock lasers}
The highly coherent radiation to interrogate the atoms is generated from diode laser systems whose frequencies are stabilized to non tunable reference Fabry-Perot resonators by means of the Pound-Drever-Hall technique \cite{dre83}.  The temperature-stabilized, vibrationally- and acoustically-isolated cavities \cite{you99}, with finesses around $10^5$, provide low noise, high-Q reference resonators for the diode laser systems \cite{oat00}.
The laser systems are based on anti-reflection-coated commercial laser diodes in an extended cavity design and provide cost-effective highly coherent sources.
With the Pound-Drever-Hall stabilization technique linewidths as low as 1~Hz (see Fig.~\ref{LDLDBeat}) can be produced. The laser frequencies are widely tunable due to the application of acousto-optic modulators (AOMs) to provide frequency offsets between laser sources and reference resonators.
Both laboratories utilize such lasers as highly stable master lasers. Several slave lasers are injection-locked to the master lasers in order to obtain the higher power (tens of milliwatts) required for the interrogation of the clock transition and the second-stage cooling.

\subsubsection{Atom interferometry in the time domain}
\label{AtomInterferometry}
\begin{figure}
\begin{minipage}[t]{\SingleFigureWidth}
\includegraphics[width=\SingleFigureWidth]{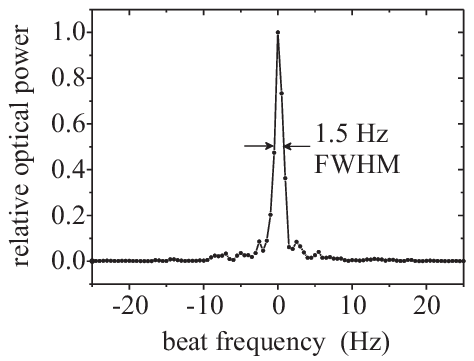}
\caption{Beat note between two diode lasers, locked to two independent reference resonators (PTB).}
\label{LDLDBeat}
\end{minipage}
\hfill
\begin{minipage}[t]{\SingleFigureWidth}
\includegraphics[width=\SingleFigureWidth]{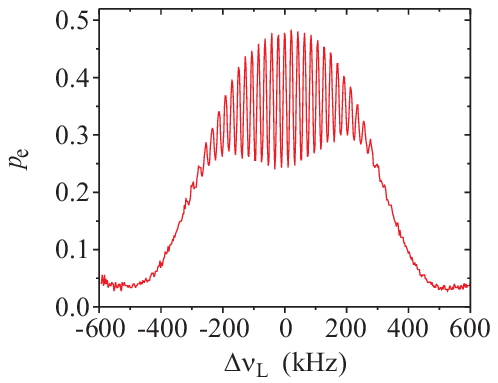}
\caption{Excitation probability for the microkelvin ensemble as function of laser detuning for the asymmetric four-pulse atom interferometer with 11.6~kHz resolution and pulse length set to $2.2~\mu$s, obtained at NIST.}
\label{ExperiRamsey}
\end{minipage}
\hfill
\begin{minipage}[t]{\SingleFigureWidth}
\includegraphics[width=\SingleFigureWidth]{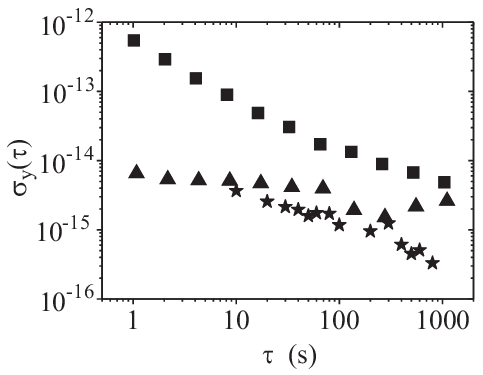}
\caption{ Combined relative Allan standard deviations. PTB: during frequency measurement of Ca-locked laser against H-maser (squares), cavity-locked laser against Ca-locked laser (triangles). NIST: lasers locked to Hg$^+$ (see \cite{ude01}) and Ca, respectively (stars).}
\label{StabilityPTBNIST}
\end{minipage}
\end{figure}

A Doppler-free method that can be described as optical Ramsey excitation \cite{bor84} or as an atom interferometer \cite{bor89} is used to overcome the limitations due to the residual Doppler broadening of about 2~MHz and 150~kHz for the millikelvin and microkelvin regimes, respectively.  In the time domain a sequence of two pairs of laser pulses from opposite directions are applied to the atomic ensemble. Each pulse provides an excitation probability of 50~\% for atoms on resonance (Rabi-angle of $\pi/2$), thereby preparing each atom in a superposition of ground and excited states, and redirecting the atomic matter waves due to the photon recoil on two closed interferometer paths of the Mach-Zehnder type. For each atom passing through either of the closed interferometers the excitation probability at the exit depends on the accumulated phase differences between the partial waves gathered along the paths:
\begin{equation}
\Phi_{1234}= 4\pi T(\nu_\mathrm{L}-\nu_\mathrm{Ca} \pm \delta_\mathrm{rec})
- \phi_\mathrm{A}(t_1,\vec{r}_1)
+ \phi_\mathrm{A}(t_2,\vec{r}_2)
- \phi_\mathrm{B}(t_3,\vec{r}_3)
+ \phi_\mathrm{B}(t_4,\vec{r}_4).
\label{PhiAsym4Pulse}
\end{equation}
The phase difference depends on the detuning of the laser frequency $\nu_\mathrm{L}$ from the clock transition $\nu_\mathrm{Ca}$ including the recoil shift $\delta_\mathrm{rec} = \hbar k^2 / 4\pi m$, with $m$ the mass of the Ca atom, and a phase contribution seen by the atom due to the instantaneous laser phases $\phi_{d}$ at the time $t_l$, and the atomic position $\vec{r}(t_l)$ of each pulse from direction $d=(\mathrm{A,B})$.
The interference pattern, i.\,e.,~the fractional excitation as function of the laser detuning, has a cosine shape and is basically symmetric with its center at the clock transition (Fig.~\ref{ExperiRamsey}) when the contribution of the instantaneous laser phases is zero. Thus, for the realization of the frequency standard, the laser can be locked to the central fringe. The slight asymmetry of the envelope of the excitation spectrum shown in Fig.~\ref{ExperiRamsey} is induced by atomic recoil in the absorption and stimulated emission processes and is an inherent effect. The position of the central fringe with respect to the frequency of the undisturbed clock transition is shifted by less than 100~mHz.

In order to achieve almost equal excitation probabilities for all atoms in the Doppler distribution, the widths $T_\mathrm{p}$ of the squares pulses are set to 1 to $2~\mu$s to achieve Fourier transform widths of up to 1~MHz. By setting the interval $T$ between the parallel pulses to several hundred microseconds the interferometer resolutions can be well below 1~kHz.

A critical aspect of this work hinges on the observation that a non-zero offset phase shifts the fringe pattern with respect to $\nu_\mathrm{Ca}$ and leads to a systematic shift in the measured transition frequency given by
\begin{equation}
\Delta\nu=-(
- \phi_\mathrm{A}(t_1,\vec{r}_1)
+ \phi_\mathrm{A}(t_2,\vec{r}_2)
- \phi_\mathrm{B}(t_3,\vec{r}_3)
+ \phi_\mathrm{B}(t_4,\vec{r}_4)
)/(4\pi T).
\end{equation}
Quantifying the contributions to this shift is an essential part of determining the uncertainty of the frequency standard.
At present, a direct laser light phase shift and an indirect shift due to the Doppler effect are the biggest contributors to the uncertainty budget (see section~\ref{Uncertainty}).

\subsection{Fractional frequency stability}
\label{StabilitySection}
A key ingredient to a frequency standard is good stability. The associated short averaging times required to reach a given frequency uncertainty greatly accelerate measurements and help to identify systematic shifts more efficiently.
Due to the large number of absorbers $N_0$, optical frequency standards with neutral atoms have the potential to provide a much better SNR and consequently a lower instability as compared to standards based on single trapped ions. The quantum-projection noise involved in the readout of the spectroscopic signal \cite{ita93} sets the limit for the minimum fractional frequency instability achievable to  $1/(\pi\cdot Q\cdot \mathrm{SNR})$, with the quality-factor $Q$ being the ratio of the transition frequency ($\nu$) to the linewidth ($\sim 1/(4T)$), and SNR$~\propto\sqrt{N_0 \tau/T_{\mathrm{cycl}}}$, where $\tau$ denotes the averaging time and $T_\mathrm{cycl}$ the cycle time.  Using the normalized detection scheme, the detection process does not limit the performance  of the standard.
This quantum projection noise limit has already been demonstrated with a neutral atom microwave clock \cite{san99}. To reach this limit in an optical clock, however, extremely narrow-band laser sources are required. Otherwise, the Dick effect \cite{dic90} will drastically limit the achievable stability. This is caused by aliasing of high-frequency laser frequency fluctuations introduced by sampling of the instantaneous laser offset from the line center in discontinuous interrogation schemes. 
For typical parameters of the PTB setup with the laser frequency noise dominated by low-frequency acoustic and seismic noise leading to a laser linewidth of 1~Hz (see Fig.\ref{LDLDBeat}) and the poor duty cycle between cooling and interrogation, the degradation of the stability due to the optical Dick effect \cite{que03} results in a relative Allan deviation \cite{all66} $\sigma_\mathrm{y}(\tau)=2 \cdot 10^{-14}~\tau^{-1/2}$ . At present a stability of $\sigma_\mathrm{y}(100~\mathrm{s})=3 \cdot 10^{-15}$ has been observed.
With a cycle time reduced to 15~ms (but using a smaller number of atoms) at NIST a fractional Allan-deviation of $\sigma_\mathrm{y}(10~\mathrm{s})=3 \cdot 10^{-15}$ has been observed, that averages down for integration times of 800 seconds to $3\cdot10^{-16}$ (see Fig. \ref{StabilityPTBNIST}).
These measurements were referenced to laser light from the Hg$^+$ single-ion frequency standard \cite{raf00} by use of a femtosecond-laser comb \cite{ma04}.
In both Ca systems the deviations from the QPN-limits are consistent with the contribution of the optical Dick effect. Even taking the low atom number at NIST and the Dick effect contribution into account, measurements of systematics (comparison of two data sets) with a relative uncertainty of $2\cdot10^{-15}$ can be achieved within 200~s.

\section{Uncertainty of the calcium optical frequency standard}
\label{Uncertainty}
In this section we present the results of frequency measurements and investigations of systematics obtained in both laboratories utilizing ensembles of millikelvin and microkelvin atoms. We discuss the various sources of the systematic shifts and their corresponding contributions to the current overall uncertainty budget. Since one of the primary goals of this research is to identify potential limitations for neutral atom frequency standards, we conclude the section with both present and projected uncertainty budgets for the Ca optical standard.

\subsection{Residual first-order Doppler shifts}
\label{ResidualFirstOrderDopplerShifts}
One significant contribution to the uncertainty of the Ca frequency standard is connected with residual atomic motion, and it is usually referred to as {\it{residual first-order Doppler effect}}.
This is to distinguish it from the relativistic {\it{second-order Doppler effect}} due to the different inertial systems of the laboratory and the moving atoms. The second-order Doppler effect is negligible for laser-cooled ensembles in the millikelvin and microkelvin regime as opposed to thermal beams. It shifts the frequencies of millikelvin and microkelvin ensembles by less than 5~mHz and $20~\mu$Hz, respectively.

The residual first-order Doppler effect is due to the ballistic movement of the atoms released from the MOT through the spatial wave fronts of the laser beams. During the time of the atom interferometry (on the order of 1~ms) each atom ``sees'' a different spatial laser phase in each interaction with one of the four laser pulses that contribute to the overall phase shift in eq.~\ref{PhiAsym4Pulse}.
Approximating the wave fronts of the two beams at the position of the atomic cloud by spherical waves with wave vectors $\vec{k}_d$  and radii of curvature $R_d$, one finds a quadratic dependence of the phase shift on the rms-width of the velocity distribution of the ensemble in a plane tangential to the curvature.  A small tilt $\beta$ of the wave vector $\vec{k}_d$ with respect to the gravitational equipotential plane leads to a shift proportional to the angle and independent of the initial velocity of the atoms. Both shifts increase with the pulse separation $T$ as $T^2$.

As has been pointed out by Trebst et al.~\cite{tre01}, one can actually use the atoms themselves to quantify and reduce these shifts. By employing different types of atom interferometers, one can evaluate the laser-beam parameters at the location of the atomic cloud. A particularly useful interferometer uses three parallel laser pulses separated by a time $T$ from a single direction. The excitation probability in this {\it{symmetric three-pulse atom interferometer}}  \cite{fri93,omi98} does not depend on the laser frequency but on the instantaneous laser phases as
$\Phi_{123}=-\phi_\mathrm{A}(t_1,\vec{r_1}) + 2\phi_\mathrm{A}(t_2,\vec{r_2}) - \phi_\mathrm{A}(t_3,\vec{r_3})$
(compare eq.~\ref{PhiAsym4Pulse}). It has the same dependence on tilt and radius of curvature as the four-pulse atom interferometer. Thus it can be used to measure the contributions for either of the two beam directions $d=\mathrm{A}$ or $d=\mathrm{B}$ (eq.~\ref{PhiAsym4Pulse}).
\begin{figure}
\begin{minipage}[t]{\SingleFigureWidth}
\includegraphics[width=\SingleFigureWidth]{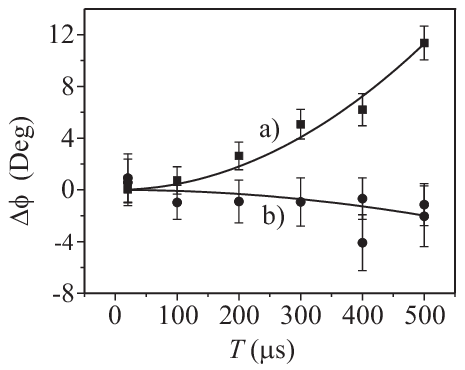}
\caption{Measured phase shifts in an interferometer due to a misaligned laser beam ($\beta = 1.6$~mrad, $R_1 \approx 12$~m) \cite{wil02c}. Measured at an ensemble of  (a) cold atoms ($\approx 2.8$~mK) and (b) ultracold atoms ($\approx 14~\mu$K)}
\label{Phasenverschiebungen}
\end{minipage}
\hfill
\begin{minipage}[t]{\SingleFigureWidth}
\includegraphics[width=\SingleFigureWidth]{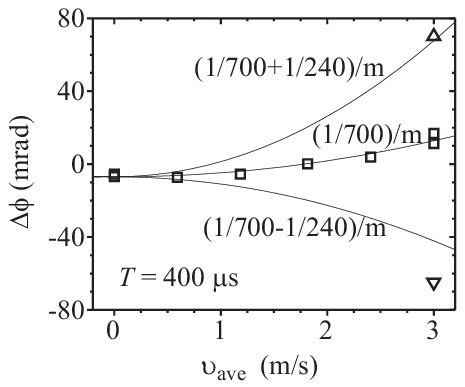}
\caption{Phase shift $\Delta\Phi$ measured as function of launching velocity. The measured shifts (points) are shown along with the estimated sensitivity (lines) for three different radii of curvature (squares, upper and lower triangles). Values given for radii are estimated from first measurement (squares). }
\label{RadiusSensitivity}
\end{minipage}
\hfill
\begin{minipage}[t]{\SingleFigureWidth}
\includegraphics[width=\SingleFigureWidth]{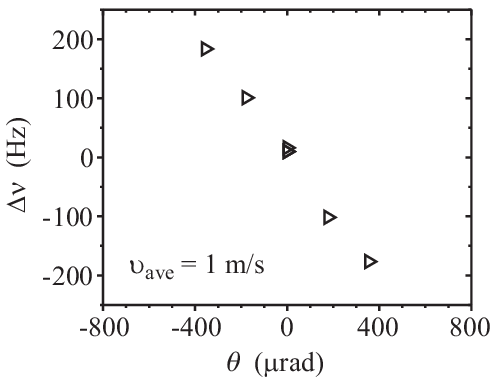}
\caption{Sensitivity of frequency shifts as function of mutual beam tilt. Launching at 1 m/s perpendicular to the beams. The points are measurements for tilting the beams along the launch direction.}
\label{TiltSensitivity}
\end{minipage}
\end{figure}
In frequency measurements performed at PTB in 2001 using 3~mK atomic ensembles, these atom interferometers were used to determine the contribution of the Doppler effect to the systematic frequency shifts, and a correction with a relative uncertainty below $10^{-14}$ was applied \cite{wil03}.
A contribution that cannot be quantified with the three-pulse atom interferometers is due to the mutual angle $\theta$ between the counterpropagating beams, resulting in a non-zero sum-vector of the two wave vectors. An average velocity of the ensemble along this sum-vector leads to a phase shift proportional to the tilt angle. Nevertheless, by carefully overlapping the counterpropagating beams the uncertainty contribution with the millikelvin ensemble was reduced to well below $10^{-14}$ \cite{wil03}.

In subsequent frequency measurements in 2003 at PTB utilizing atoms at a temperature of  $12~\mu$K, the influence of the velocity-dependent wave-front curvature was drastically reduced. Fig.~\ref{Phasenverschiebungen} shows this for the example of a particularly high wave-front curvature. Even with the very large radius of 12~m the contribution to the overall Doppler effect is less than $10^{-15}$ for the microkelvin ensemble. Consequently one is left with only the contribution from the angle to gravity $\beta$. In measurements at PTB, the symmetric three-pulse atom interferometer was used to align $\beta$ to within $100~\mu$rad.  This helped to reduce the total uncertainty level due to the residual Doppler effect to $2\cdot10^{-15}$.

At NIST, this approach has been taken a step further by launching the clouds of atoms in order to enhance the sensitivity of the atom interferometers. Additionally, launching allows one to evaluate the various beam parameters independently, which enables the following straightforward optimization procedure.
One first reduces the curvature of the wavefronts by using the three-pulse interferometer with atomic clouds launched up to 3~{m/s}. Due to the quadratic sensitivity of the phase on velocity, one can readily measure radii of curvature $R_d$ of thousands of meters (see Fig.~\ref{RadiusSensitivity}) with NIST's present signal-to-noise ratio.
With the effects due to curved wavefronts thus reduced, subsequent measurements using the four-pulse interferometer can then enable reduction of the overlap angle $\theta$ to less than $1~\mu$rad  (Fig.~\ref{TiltSensitivity}), corresponding to a frequency shift of 0.7~Hz at a launch velocity of 1~m/s.
Finally, the contribution of the tilt angle can be largely suppressed with a beam-reversal technique due to the minimized radius of curvature and overlap angle contributions.
In principle this sequence allows one to reduce the Doppler-shift contribution to the fractional frequency uncertainty to below $10^{-17}$, thereby virtually removing one of the most challenging systematic effects for the standard.  However, detailed measurements have revealed that the present laser-beam quality prevents performance below a level of $10^{-15}$. While the sensitivity of the method and the prospects for laser beams of higher quality are certainly good, the issue of beam quality clearly remains one of the most critical for the future performance of the standard.

\subsection{Laser light phase shifts}
\label{LaserLightPhaseShifts}

In addition to the influence of the Doppler effect on the resulting phase shifts, a direct shift of the global phase of the laser beams has to be taken into account.
As the groups at NIST and PTB have investigated, such a phase shift can occur from a phase chirp produced in the switching AOMs \cite{ude01,cur03a,deg04x}. A ringing on the edges of the RF pulses (and of the acoustic wave in the AOM) is imprinted on the laser pulses.  This leads to a constant total offset phase and, hence, to a frequency shift proportional to the resolution in the four-pulse atom interferometer, i.\,e.,~proportional to $T^{-1}$. By measuring the frequency shift as a function of resolution and extrapolating to zero fringe linewidth, one can correct for this effect. Consistent results using this correction have been achieved at NIST and PTB at respective levels of $10^{-14}$ and $3\cdot10^{-15}$. Clearly, reduction of these uncertainties emerges as one of the most important issues for future frequency measurements. The determination and correction of these effects, on the other hand, will be limited ultimately by the remaining uncertainty of the residual first order Doppler effect, since both effects depend on the pulse separation $T$.

\subsection{Frequency shifts due to external fields}
External fields interacting with the atoms can shift the frequency of the atoms during interrogation with respect to unperturbed atoms. On the one hand, constant or slowly fluctuating magnetic and electric dc fields shift the frequency due to magnetic effects and the quadratic Stark effect. On the other hand, thermal radiation from the environment and the oven as well as monochromatic radiation from the lasers can shift the frequency due to the ac-Stark effect.

\subsubsection{Magnetic field shifts}
The $^1$S$_0$\,($m=0$) to $^3$P$_1$\,($m=0$) clock transition does not have a linear Zeeman shift and has only a small second-order Zeeman shift. The value of the second-order Zeeman shift of the $^3$P$_1$ state is known from theoretical calculations \cite{bev98} to be $(+63.75 \pm 0.09)$~Hz/mT$^2$.  The diamagnetic shift of the $^1$S$_0$ state is likely to be below 1~\% of this value.
Measurements on atomic beams and laser-cooled atoms \cite{bev87a,zin98,oat99} show the second-order Zeeman shift of the clock transition to be $(+64 \pm 1)$~Hz/mT$^2$.
Typically a quantization field of 0.1 to 0.2~mT separates the Zeeman components ($\Delta m=\pm1$) well from the $\Delta m=0$ component. This field can be measured by use of the Zeeman components that are separated from the $\Delta m=0$ component with $\pm 21.01$~MHz/mT \cite{bev87a}. When measuring the field with a relative uncertainty of 0.5~\%  and using the theoretical coefficient with an uncertainty of 1~\%, the quadratic Zeeman shift at 0.2~mT can be corrected with an uncertainty of 36~mHz.

\subsubsection{DC electric fields}
The coefficient for the quadratic Stark effect is well known from measurements by Zeiske \cite{zei95a} of the differences $\Delta\alpha$  of the static polarizabilities of the clock states. For electric stray fields stemming from arbitrary electric charges on the surfaces of the vacuum chambers and magnetic field coils (distance to trap $\geq 3$~cm) an average value $\overline{\Delta\alpha}=(3.07\pm0.06)~\mu$Hz/(V/m)$^2$ can be used to correct for shifts. If we assume that electric potentials are less than 10~V on the grounded surfaces ($E \leq 200~$V/m), the resulting shift lies between 0 mHz and -70~mHz. If we assume equal probability of a shift within this interval, the average shift is $(-35 \pm 20)$~mHz. Higher-order contributions shift the transition by not more than a few microhertz for dc field levels up to 1~kV/m.

\subsubsection{Blackbody radiation}
\label{BlackbodyRadiation}
The influence of the radiation due to the ambient temperature in the lab (between 288~K and 298~K) can be estimated based on a quasi-static approach \cite{zei95a}.  With the atoms placed in a blackbody-radiation environment of temperature $\vartheta$ the systematic shift is $(-1.31\pm 0.03)\cdot 10^{-10}~\frac{\mbox{Hz}}{\mbox{K}^4}\cdot\vartheta^4$.
The influence of the radiation of the Ca oven at 900~K can be obtained from an exact calculation based on an approach by Farley and Wing \cite{far81}, which includes all relevant transitions. The calculated shift is about 45 \% larger than the one found with the quasi-static approach.
However, it is difficult to estimate the level of radiation from the oven reaching the atomic cloud. Therefore measurements of the shifts at various oven temperatures were performed at PTB during the frequency measurements in 2001 and 2003, resulting in an estimate of the shift of $(-2.8\pm 3.9)$~Hz. This is at present the largest contribution to the total uncertainty.
Proper shielding of the trap from the oven, possible with the use of a Zeeman slower and an optically deflected atomic beam, will allow one to eliminate the oven contribution. This will reduce the uncertainty in the correction of the blackbody shift to the contribution of the room-temperature. For a temperature of $\vartheta = 293$~K known to $\pm1$~K the uncertainty in the correction could then be reduced to 24~mHz.

\subsubsection{AC Stark effect}
\label{ACStarkEffect}
In the PTB setup the influence of residual laser light from the cooling and quenching lasers is eliminated by the use of mechanical shutters.  At NIST the light is switched off using multiple deflections in AOMs. Measurements with increased residual light levels, when AOMs were left on, showed that even when the frequency offsets with the AOMs switched off are neglected, the shifts are less than 1~mHz for the blue cooling laser and 85~mHz for the quench laser.
Presently, the red light in the cooling and probing beams is less well isolated and can lead to frequency shifts of up to 0.5~Hz. Cascading multiple AOMs in a suitable way will allow one to reduce this shift to the millihertz level as well.
Hence, even at a relative uncertainty level of $10^{-17}$, the use of mechanical shutters can be avoided, thereby avoiding compromises regarding cycle times and stability.

\subsection{Collisions with thermal atoms}
\begin{figure}
\begin{minipage}[t]{\SingleFigureWidth}
\includegraphics[width=\SingleFigureWidth]{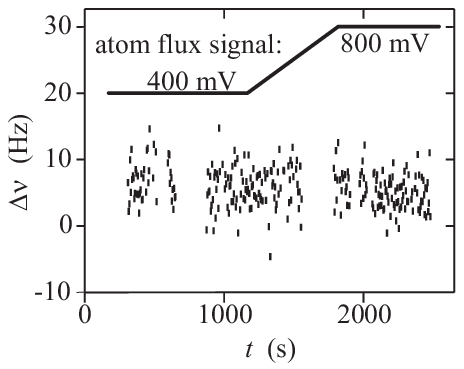}
\caption{Measurement of the change of the Ca clock frequency with a changing flux of thermal atoms. Accessed via NIST's femtosecond-laser comb, the mercury single-ion frequency standard was used as a reference. The difference found for doubling the atomic flux is $(-0.9\pm2.2)$~Hz.}
\label{ThermalCollisions}
\end{minipage}
\hfill
\begin{minipage}[t]{\DoubleFigureWidth}
\begin{center}
\includegraphics[width=\DoubleFigureWidth]{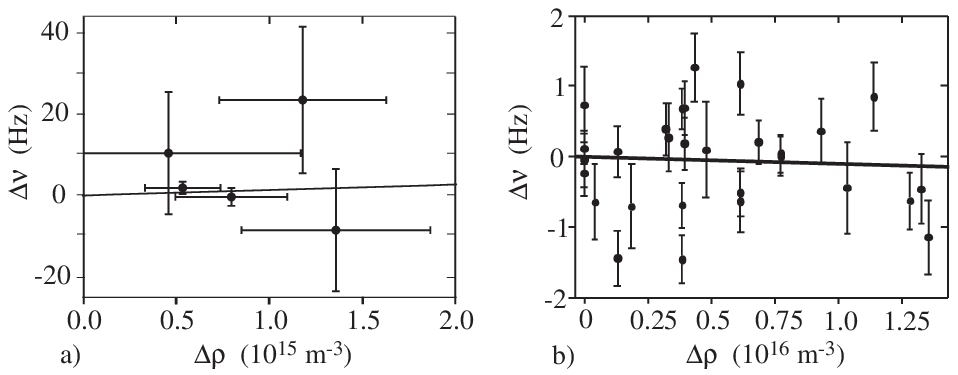}
\caption{Measurement of the density-dependent frequency shift {\bf (a)} using cold atoms ($T \approx$~3~mK), {\bf (b)} using ultracold atoms ($T \approx 20~\mu$K).
    The lines represent a linear regression to the data}
\end{center}
\label{Stoesse}
\end{minipage}
\end{figure}

In the PTB setup the cold ensemble is shielded from collisions with thermal atoms by a beam block placed in the path of the thermal beam. Without the aperture, trap losses due to collisions with thermal atoms would strongly reduce the number of atoms in the trap.
In the similar NIST setup no aperture is used, but the oven is run at a lower temperature. The influence of collisions with thermal atoms on the frequency shift was measured referencing the Ca frequency to NIST's optical Hg$^+$ frequency standard and to NIST's H masers using the femtosecond-laser comb. Fig.~\ref{ThermalCollisions} shows a measurement against Hg$^+$ when the flux of atoms was changed by a factor of 2. No frequency change was observed that would be significant on a 1~Hz level. In the future it can be completely avoided by using a deflected slow atomic beam for loading the MOT.

\subsection{Influence of cold and ultracold atomic collisions}
\label{ColdCollisions}
Interactions between the atoms in the laser-cooled ensemble, i.\,e.,~cold collisions, lead to a frequency shift of the clock transition. The magnitude of the shift depends on the states of the two colliding atoms and their separation, and results in a mean frequency shift that depends on the temperature and the density of the atomic ensemble. For the low densities ( about $10^{16}$\ m$^{-3}$) in PTB's ensemble, the frequency shift is assumed to be a linear function of the collision rate, and hence of the density $\rho$.
In order to determine the density dependence of the frequency shift, the frequency differences were measured as a function of average density differences \cite{wil03}. The density was varied by switching the repump laser at 672~nm (Fig. \ref{LevelSchemes}) on or off during the loading time of the millikelvin ensemble. This changed the number of trapped atoms while all other parameters such as the magnetic field, loading time, frequency of the trapping light and laser-beam profile were kept constant. The radius and temperature of the trapped cloud were checked and did not change significantly when the average density $\Delta \rho$ was varied by a factor of five. The measurement of the collisional shift $\Delta \nu / \nu$ depending on the density $\rho$ (Fig. \ref{Stoesse}~a) gave $\Delta \nu / \nu = (3 \pm 4.4) \cdot 10^{-30}~\mathrm{m^3}\times \rho$~\cite{wil03}.  The corresponding frequency shift for the average density of the expanding ensemble when measuring at a resolution of 1.2~kHz ($T=215~\mu$s) would be $(1.2\pm1.8)$~Hz. Hence, within the uncertainty of this measurement, no significant shift has been observed.
The same was true when microkelvin atoms were used (Fig. \ref{Stoesse}b) where an even lower limit of $\Delta \nu / \nu = (-0.4 \pm 1.2) \cdot 10^{-31}~\mathrm{m^3}\times \rho$ was obtained. At the usual densities at the PTB setup this would lead to a shift of $(20 \pm 60)$~mHz, while in the NIST setup the densities and therefore also the shifts are lower by at least one order of magnitude.
Within its uncertainty this coefficient is four orders of magnitude smaller than that seen in Cs-fountain clocks \cite{per02}, which have nonetheless achieved an uncertainty level of $10^{-15}$. Thus one sees the potential of this optical frequency standard to achieve very low uncertainty levels while maintaining a very high stability with a high number of atoms ($\sim 10^7$).

\subsection{Frequency measurements and uncertainty budget}
\begin{figure}
\begin{center}
\includegraphics[width=110mm]{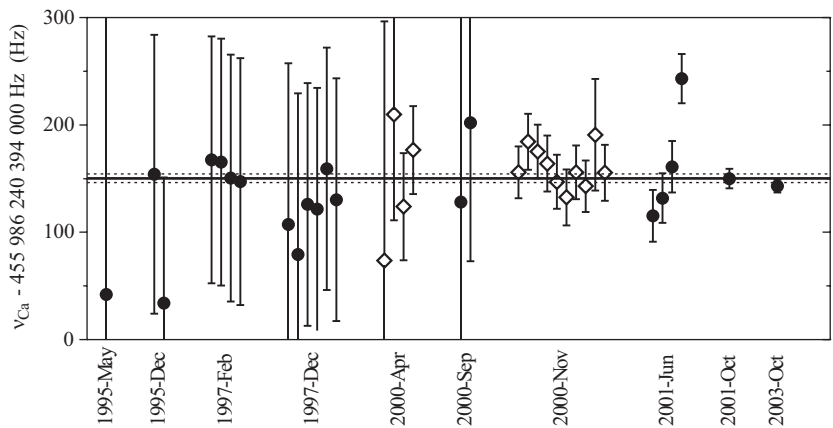}
\caption{Results of frequency measurements of the 657~nm Ca intercombination line performed at PTB (solid circles) and NIST (open diamonds) with the weighted average (solid line) of all measurements and the statistical uncertainty (dashed lines).}
\label{NuCas}
\end{center}
\end{figure}

Table~\ref{TabUncertainty} summarizes the results for the present uncertainty budget for frequency measurements with millikelvin atoms in 2001 at PTB \cite{wil02,hel03} and 2000 at NIST \cite{cur03a,vog01,ude01}, as well as a recent frequency measurement with microkelvin atoms in 2003 at PTB.
\begin{table}
\centerline{
\begin{tabular}{c|ccc|c}
\hline
                        & \multicolumn{3}{c|}{Uncertainty (Hz)}                 & Projected Uncertainty (mHz)\\
\hline
                        & NIST      & PTB               & PTB           &           \\
                        & Oct./Nov. 2000~~  & Oct. 2001         & Oct. 2003         &           \\
Effect                          & ($T = 3$~mK)      & ($T = 3$~mK)      &  ($T = 12~\mu$K)  &   ($T = 12~\mu$K) \\
\hline
1st-order Doppler effect                    & 12.1          & 2.6           & 1.0           &   4           \\
2nd-order Doppler effect                    & 0.005         & 0.005         & $2\cdot10^{-5}$   &   0.02        \\
other phase contributions                       & 4         & 4             & 1.6           &   5           \\
asymmetry of line shape                         & 0.05          & 0.05          & 0.05          &   5           \\
2nd-order Zeeman effect         & 1             & 0.1           & 0.1           &   36          \\
AC-Stark effect                 & 2             & 0.1           & 0.1           &   1           \\
quadratic Stark effect              & 0.06          & 0.02          & 0.02          &   20          \\
blackbody radiation from            &           &           &           &           \\
\hspace{0.5cm}oven                      & 1         & 4.3           & 3.9           &   --          \\
\hspace{0.5cm}chamber           & 0.07          & 0.07          & 0.07          &   24          \\
collisions of cold atoms            & 10            & 1.8           & 0.06          &  10           \\
electronic stabilization \& laser drift     & 10            & 3.2           & 0.1           &  1            \\
\hline
sum in quadrature               & 19.2          & 7.4           & 4.3           &           \\
\hline
statistical uncertainty of frequency meas.  & 2.5           & 3         & 3         &   0.05        \\
Cs clock (about $1 \cdot 10^{-15}$)                 & 0.8           & 0.5               & 0.5           &           \\
\hline
total uncertainty               & 19.4          & 8             & 5.3           &           \\
\hline
\bf{total relative uncertainty}
    $\mathbf{\delta \nu / \nu}$             & $\mathbf{4.3 \cdot 10^{-14}}$
                                    &$\mathbf{1.8 \cdot 10^{-14}}$
                                                & $\mathbf{1.2 \cdot 10^{-14}}$
                                                            &
                                                                        \\
\hline
\end{tabular}
}
\caption{Uncertainty budget for frequency measurements utilizing the Ca frequency standards at PTB and NIST  with millikelvin atoms (col. 1, 2) \cite{wil02c,cur03a}, and  with microkelvin atoms (col. 3). Column~5 summarizes the projected uncertainty budget for an optimized Ca frequency standard (see text).}
\label{TabUncertainty}
\end{table}
While the measurements at NIST in 2000 were still limited by the unknown effect of cold collisions and the exact contributions by the Doppler effect, the PTB measurements of 2001 were limited by the blackbody shift (4.3~Hz) due to the oven being close to the atomic ensemble.  This contribution still dominates the uncertainty budget in PTB's frequency measurement of 2003, while the Doppler effect and subsequently other phase contributions were further reduced by use of microkelvin atomic ensembles.  A setup with the oven removed from the vicinity of the trapped ensemble is currently being assembled.
Asymmetries in the excitation spectra that can lead to systematic frequency shifts when stabilizing the laser have been estimated as well as measured \cite{zin98} indicating a shift of less than 50~mHz with a 3f-stabilization scheme.
The use of femtosecond-laser frequency combs for the comparison against state-of-the-art Cs-fountain atomic clocks (and against the Hg$^+$ single-ion frequency standard at NIST) allowed both groups to avoid being limited by either the primary time and frequency standards or the statistical uncertainty of the comparison.
The results are summarized in Fig.~\ref{NuCas} together with the first frequency measurements of the Ca clock transition performed at the PTB, using phase-coherent frequency multiplication chains \cite{sch96,zin98}.
The average over all measurements is ${\bar{\nu}_\mathrm{Ca}} =  (455\,986\,240\,494\,150.2 \pm 7.7)$~Hz.
%
%
The latest measurement with microkelvin atoms performed at PTB in October 2003 resulted in $\nu_\mathrm{Ca}= (455\,986\,240\,494\,144 \pm 5.3)$~Hz.
The results of all measurements agree well with the average value within their respective uncertainties (see fig. \ref{NuCas}). This makes the Ca optical frequency standard with neutral atoms the best such standard evaluated in more than one laboratory.

\subsection{Projected uncertainty budget}
\label{ProjectedUncertaintyBudget}
The projected uncertainty budget in the last column of Table~\ref{TabUncertainty} is based on the present knowledge on the system as presented in the above sections.
Possibly, the dominating uncertainty contribution will result from the correction needed for the room-temperature blackbody shift.  The limit is due to the uncertainty in the correction.
Modelling of the atomic signal (see \cite{wil02c}) based on a description by Bord\'e et al.~\cite{bor84} to correct for shifts due to the asymmetries in the line shape (see Fig.~\ref{ExperiRamsey}) will allow one to quantify the shift more accurately. A correction of the shift to within 2~\% would reduce the uncertainty to 1~mHz.

Electronic stabilization to line center at a level better than $10^{-6}$ has already been demonstrated in the NIST-7 thermal atomic Cs-clock \cite{lee95}. An offset from the line center due to drift of up to 10~Hz/s of the free-running laser can be reduced to below one millihertz when two integrator stages and measurement cycles on the 10~ms range are used (see \cite{zin98} for a mathematical model describing such a stabilization scheme).

With $10^7$ atoms confined to a trap of radius 2~mm the present knowledge on density shifts would be sufficient to make this uncertainty contribution negligible.
Furthermore, local higher-order distortions of wave fronts on a scale of 1~mm that cannot be clearly resolved by launching the ensemble would be averaged away over the size of the trap. A 12~$\mu$K ensemble with an average velocity of less than 1~mm/s would enable one to achieve a negligible residual first-order Doppler effect. Nevertheless a wave-front flatness with $R_d > 3000$~m would need to be achieved simultaneously and would certainly present a technical challenge.

The numbers given in the table are best-case estimates based on our present knowledge of what could be achieved. Whether all these conditions can be simultaneously achieved, as would be required for a total combined uncertainty below $10^{-16}$, is subject to further investigation.

\section{Conclusion}
We have described the current status of the optical Ca frequency standard with laser-cooled neutral atoms as realized in two different laboratories for the purpose of developing a possible future optical atomic clock.
Frequency measurements performed make the frequency of the clock transition of $^{40}$Ca one of the best known optical frequencies (uncertainty $1.2\cdot10^{-14}$), and agree in both laboratories to well within their respective uncertainties.
Furthermore, extensive investigations of systematics carried out in both laboratories do indicate that the presently visible roadblocks are of a purely technical nature and that measurement capabilities to identify these shifts already exist. With the present knowledge of the properties of this frequency standard an attainable uncertainty level of below $10^{-16}$ can be anticipated.

Taking advantage of the unprecedented potential stability of $10^{-16}$ in 1~s that is feasible with this standard, the systematic uncertainty level could be reached in a few seconds of measurement time. This would however require a free running spectroscopy laser of low frequency noise during the 100~ms needed for a full stabilization cycle. In case of white frequency noise this would correspond to a laser linewidth of less than 100~mHz.

Such a standard is a promising candidate not only for an application as a future atomic clock, but as well as a versatile instrument for precision measurements including the search for variation of the fundamental constants and the test of fundamental theories (e.\,g.\ see \cite{pei04} and references therein).
%



\section*{Acknowledgements}
The {PTB} group acknowledges support from the Deutsche Forschungsgemeinschaft under {SFB~407} and the European Union under the Research and Training Network {CAUAC}. The {NIST} group acknowledges support in part by {NASA} and {OMR-MURI}. G. Wilpers acknowledges support by the Alexander von Humboldt-Foundation. The NIST group is grateful for the experimental support with the Hg$^+$ frequency standard by Windell Oskay and Jim Bergquist, with the femtosecond-laser comb by Albrecht Bartels and Scott Diddams, and with the {NIST} time scale by Steve Jefferts, Tom Heavner, Elizabeth Donley and Tom Parker.


\end{document}